\documentclass[12pt]{iopart}
\usepackage{graphicx,color,amssymb,amsfonts,hyperref}
\usepackage[usenames,dvipsnames]{xcolor}
\usepackage[frenchb,english]{babel}
\definecolor{nblue}{rgb}{0.3,0.3,1.0}
\definecolor{ngreen}{rgb}{0.2,0.7,0.2}
\definecolor{nred}{rgb}{0.9,0.1,0}
\definecolor{npurple}{rgb}{0.8,0.2,0.8}
\definecolor{golden}{rgb}{0.8,0.6,0.1}
\definecolor{nsilver}{rgb}{0.3,0.4,0.5}
\definecolor{nbrown}{rgb}{0.8,0.4,0.15}
\definecolor{nrose}{rgb}{0.7,0,0.35}
\definecolor{nviol}{rgb}{0.5,0,1.0}
\definecolor{nazur}{rgb}{0,0.35,0.7}
\definecolor{nchart}{rgb}{0.2,0.4,0}

\newcommand{\blk}{\color{black}}
\newcommand{\blu}{\color{nblue}}
\newcommand{\grn}{\color{ngreen}}
\newcommand{\red}{\color{nred}}

\newcommand{\gold}{\color{golden}}

\newcommand{\beq}{\begin{equation}}
\newcommand{\eeq}{\end{equation}}
\newcommand{\bqa}{\begin{eqnarray}}
\newcommand{\eqa}{\end{eqnarray}}

\newcommand{\ie}{{\em i.e.}}

\begin{document}

\title[]{Reply to Norsen's paper ``Are there really two different Bell's theorems?'' }
\author{Howard M. Wiseman}
\address{Centre for Quantum Dynamics, Griffith University, Brisbane, Queensland 4111, Australia}
\author{Eleanor G. Rieffel}
\address{QuAIL, NASA Ames Research Center, Moffett Field, CA 94025}

\begin{abstract}
Yes. That is my  polemical reply to the titular question in Travis Norsen's self-styled ``polemical response to Howard Wiseman's recent paper.''  
Less polemically, I am pleased to see that on two of my positions ---  that Bell's 1964 theorem 
is different from Bell's 1976 theorem, and that the former does not include Bell's one-paragraph heuristic 
presentation of the EPR argument --- Norsen has made significant concessions.
 In his response, Norsen admits that  ``Bell's recapitulation of the EPR argument in [the relevant] paragraph  leaves something to be desired,'' that it ``disappoints'' and is ``problematic''. Moreover, Norsen makes other statements that imply, on the face of it, that he should have 
no objections to the title of my recent paper (``The Two Bell's Theorems of John Bell'').  My principle aim in writing that paper was to try to bridge the gap between two interpretational camps, whom I call 
`operationalists' and `realists', by pointing out that they use the phrase ``Bell's theorem'' to mean different things: his 1964 theorem 
(assuming locality and determinism) and his 1976 theorem (assuming local causality), respectively. Thus, it is heartening that at least one person from one side has taken one step on my bridge. 
That said, there are several issues of contention with Norsen, 
which we (the two authors) address after discussing the extent of our agreement with Norsen.  
The most significant issues  are: 
the indefiniteness of the word `locality' prior to 1964; and
the assumptions Einstein made in the paper quoted by Bell in 1964 and their 
relation to Bell's theorem. 
\end{abstract}

\section{Introduction} \label{sec:intro}

We  assume that the reader has read Norsen's paper in this issue~\cite{Nor15}, and hope that the 
reader has also read my paper~\cite{Wis14b} to which it was written in 
response\footnote{Despite having two authors, this paper often employs  the first 
person singular. This seemed the simplest option since the paper refers so often to the earlier 
paper by the first author (Wiseman, the first person singular), and what Norsen said about it. 
The second author (Rieffel) helped shape my understanding of the role of Bell's EPR paragraph 
in the course of correspondence as I was writing that first paper.  
Thus, it was natural for me to invite her 
to help craft  this response to Norsen, since this 
paragraph is at  the crux of our  arguments with Norsen.}.
The purpose of my paper was to obviate the heated and fruitless debates that I have observed between operationalists 
and realists that occur because they use the phrase ``Bell's theorem" to mean different theorems proven by Bell. Operationalists 
(who form probably the majority of quantum physicists, especially quantum information theorists) 
refer to Bell's most famous theorem, from 1964, proving the incompatibility  of quantum phenomena with the dual assumptions  of locality 
and determinism.
Realists (Norsen included) appeal rather to his 1976 theorem, proving   
the incompatibility  of quantum phenomena with the unitary property of local causality. The situation is confused because of 
Bell's own later assertion, which the realist camp endorses,  that his 1964 theorem began with that single, 
and indivisible, assumption of local causality 
(even if not by that name).

 Norsen's reaction to my paper 
centres on the ``EPR paragraph'' in Bell's 
1964 paper, which follows Bell's introduction/abstract. 
 There is actually less disagreement  
between us than one might think from Norsen's abstract.  
Specifically,  some of the flaws
in Bell's argument in that paragraph, from `locality' to determinism,
are now acknowledged by Norsen. He even states that he would not object to
somebody like me reserving the word `theorem' for what I called Bell's 1964 theorem. 
We  discuss these points in detail in Sec.~2, 
and then move on to some matters where  Norsen still disagrees with us. 
In Sec.~3, we  discuss how it is wrong to imply (as Norsen does) 
that the term `locality' has always had an unambiguous meaning, 
that of `local causality' (Bell's 1976 term), 
and that this meaning is to be found 
in the 1949 quote from Einstein which Bell used in 1964. 
In Sec.~4, we  discuss in more detail Einstein's 1949 argument, 
and how Norsen misrepresents both it, and my statements on it.  
In particular, we take the opportunity to clarify the status of 
``orthodox quantum mechanics" (as raised by Norsen) 
with regard to Einstein's `no-telepathy' notion  
and Bell's 1964 `locality' notion. Finally, in Sec.~5, we reflect on the bigger picture: 
why do Norsen and his realist colleagues care so much about this issue, and 
what would it mean for Bell if one were to accept their arguments?

\section{What, in Bell's 1964 paper, is his theorem?}

Norsen and the `realist' camp argue as follows regarding Bell's 1964 paper: 
\red ``First, they see the first paragraph of Bell's `Formulation' 
section as an essential part of his 1964 theorem, the first part 
of a two-part argument. Second, they see 
this first part as legitimately deriving determinism from predictability and the assumption of locality, 
since \ldots [b]asically, they see Bell's notion of locality as being local causality.'' \blk 
\blk Here we \blk follow Norsen in using \red red \blk for quotes from my own paper (although here 
expressing Norsen's views, not mine), 
\blu blue \blk for Bell's 1964 paper, and \grn green \blk for Einstein's scientific autobiographical notes  
published in 1949 (though written a few years earlier)~\cite{Ein49}. 
In  a spirit of generosity, we \blk have added \gold gold \blk for Norsen's paper~\cite{Nor15}. 

In my paper I disputed both the `First' and `Second' parts of the above reading of Bell's 1964 paper. 
 Thus, my reading was that the section Bell 
entitles `Formulation' is, indeed, just a motivation 
and formulation of the two assumptions (locality and determinism) 
needed for his following theorem. Norsen says (third paragraph of his Sec.~4) 
that this reading is \gold ``completely and utterly implausible'' \blk and that 
 \gold ``Wiseman is \ldots clearly engaging in some pretty 
creative interpretation''.  \blk  
We are not sure what Norsen 
 thinks is so creative about my interpretation, since 
 it  is --- as Norsen agrees (opening paragraph of his Sec.~2) --- the one ``almost 
 universally reported''\footnote{The words here are Bell's~\cite{Bel81}, 
and here  (in 1981) Bell was complaining that this almost universal 
reporting was a misunderstanding. 
In Norsen's view (opening paragraph of his Sec.~4) this is an \gold ``important piece of evidence'' \blk 
that it is indeed 
a misunderstanding. To the contrary, we 
think any paper, especially a contentious one, has to be 
interpreted as it appeared, rather than how the author 
would, almost two decades later, have us interpret it, even for an author with 
as high a reputation for integrity as Bell has.}. I raise this to be clear that I did  
not propose any new interpretation, but rather tried to explain as clearly as possibly to each camp what the 
other camp means by ``Bell's theorem'' and how, in both cases,  this is reasonable. 

\subsection{Disputing the \red `First' \blk point}

I  disputed the \red `First' \blk point above using many pieces of evidence 
(third paragraph of Sec.~3.2):
\red \begin{quote} 
i) it does not explain why Bell would, in 1964,  state his result {\em four times} as requiring two assumptions, 
locality and determinism, and not once as requiring only the assumption of locality; 
ii) it does not explain why in his first subsequent paper on the topic of hidden variables~\blk\cite{Bel71}\red, 
after seven years to think about how best to explain his result, he still states it 
(somewhat redundantly) as being ``that no local deterministic hidden-variable theory 
can reproduce all the experimental predictions of quantum mechanics''~\blk\cite{Bel71}\red; \ldots\ 
iv) it does not explain why Bell would state the conclusion of the supposedly crucial first part of his theorem as
being merely that it ``implies the {\em possibility} of a more complete specification of the state.''; 
v) it does not explain why Bell would place this supposedly crucial first part {\em prior} to the mathematical formulation 
of his result, and not mention it anywhere else in the paper. 
\end{quote}\blk
 In his response to my paper, Norsen does not address items (ii) and (iv),  
and, when addressing items (i) and (v), makes no arguments that I had not already 
rebutted in my paper.  Note that we  omit item (iii) in 
the list above 
not because Norsen succeeds in refuting it, but because it relates 
to the  disputation of the \red `Second' \blk  
point (see below), rather than the \red `First'. \blk

Further evidence in support of the above can be found in Ref.~\cite{Bel71}, which is, as noted above, 
Bell's first statement on the subject after his 1964 paper. In this paper it is even more obvious that the EPR argument is merely a motivation for the hidden variables assumption. Bell in fact lists three ``motivations'' for this assumption in his opening section (entitled, unambiguously, ``Motivations'') of which an EPR-style argument is the third. Unlike in his 1964 presentation, his 1971 EPR paragraph does not make an argument from locality (of any kind, or under any name) to hidden variables. Rather, he just appeals to the intuition of the reader (our emphasis): ``some quantum mechanical predictions \ldots\ {\em seem almost to cry out} for a hidden variable interpretation''; ``This [prediction by a distant party] {\em strongly suggests} that the outcomes of such measurements \ldots\  are actually determined in advance.'' It is only after making such appeals that he introduces ``locality'' as being a favourable property potentially offered by such hidden variable theories. As noted above, Bell then states his theorem as ruling out ``local deterministic hidden variable[s]'', 
which tells against any reading that has him use the EPR argument (here, or in 1964) as the first part of a two-part theorem using only the assumption of locality. 

 \blk

\subsection{Disputing the \red `Second' \blk point}

I disputed the \red `Second' \blk point above on the grounds that in 
this EPR-paragraph, prior to the formulation of his theorem,  
\red 
\begin{quotation}  ``Bell has made a mistake. His conclusion (predetermined results) does not 
follow from his premises (predictability, and the hypothesis \blk[\ie\ `locality'] \red stated in the preceding sentence.) \ldots

 It would have been an easy mistake for Bell to have made, if he had the idea 
that EPR had already proven determinism from some sort of locality
assumption, and did not think hard about whether it was the same 
as the locality assumption he was about to use in his own theorem.'' 
\end{quotation}\blk
(This is from the final paragraphs of my Sec.~3.1.)

In his paper (see the latter part of his Sec.~3), Norsen has essentially conceded this second point. 
He says  that \gold 
``Bell disappoints us. \blk \ldots \gold what he says here about locality (\blu``if the two measurements are made at places remote from one another the orientation of one magnet does not influence the result obtained with the other''\gold) certainly falls short of a general formulation 
(along the lines that he would later give, in 1976 and 1990) [of locality].'' \blk Here by a \gold 
``general formulation of locality'', \blk Norsen means what I called, 
using the term introduced by Bell in 1976~\cite{Bel76} and promoted unequivocally by him in 1990~\cite{Bel90b}, ``local causality''. 
This is the simplest assumption needed for the EPR argument from predictability to predetermined results to work. And indeed Norsen 
goes on to say \gold  ``[W]hat Bell says here seems problematic \ldots. It is simply not clear how to translate Bell's words here (about locality) 
into a sharp mathematical statement in terms of which the EPR argument might be rigorously rehearsed. \ldots\ 
So, and especially taking into account the five decades of controversy that have followed, it must be admitted that Bell's recapitulation of the EPR argument in this paragraph leaves something to be desired.'' \blk  

In the penultimate paragraph of his paper, Norsen says: 
\gold \begin{quotation}
``If a commentator wants to reserve the word `theorem' for demonstrations meeting some minimal threshold of rigor (and chooses to place the threshold somewhere [below] what Bell did after the first paragraph of \blu ``2 Formulation'' \gold in his 1964 paper) I would have no objection, so long as the commentator articulates clearly that ``Bell's [1964] theorem'', when combined with the earlier ``EPR/Einstein non-theorem'' establishing the need for deterministic hidden variables, leads to the overall conclusion that the QM predictions are incompatible with locality \ldots\ and that this is what Bell took himself to have established already in 1964. I would even have no objection if such a commentator raised questions about whether this incompatibility was really established in 1964, since (the commentator might plausibly argue) genuinely establishing such a conclusion requires that all parts of the argument leading to it meet the commentator's threshold for theoremhood.'' 
\end{quotation}\blk 
I claim to have been  just such a non-objectionable commentator!  Apart from the fact that we 
would use ``local causality'' in place of `locality' in the above,  we  see no incompatibility with the above and what I said in my paper  --- in particular with what I said 
in the penultimate paragraph of Sec.~3.1: 
\red\begin{quotation}
Bell's first paragraph (not counting the abstract) \blk [\ie~his EPR paragraph] \red 
 serves only to motivate the formulation of the theorem. \ldots\ [T]hat Bell's EPR paragraph forms no part of his 1964 theorem 
 (which in this paper he calls his \blu `result'\red) 
is clear from the fact that after he has {\em formulated} the problem 
in Sec.~II, and {\em illustrated} it in Sec.~III, he finally gets to Sec.~IV where 
\blu``The main result will now be proved.'' 
\end{quotation}\blk 
\blk and in the final paragraph of Sec.~3.2: \red
\red\begin{quotation}
Of course one is  free to  construct a proof using the EPR paragraph of Bell's 
paper to derive locality and determinism,  if one corrects it by beginning with a sufficiently strong localistic 
hypothesis \blk[such as local causality]. \red  This may be useful to obtain a pedagogical proof of Bell's theorem (in a general sense), as 
in Ref.~\blk\cite{Nor11} \red for example, but that should not be confused with  Bell's 1964 theorem, or its proof. 
\end{quotation}\blk 
Finally, I was clear that Bell represented his 1964 paper as doing exactly this, giving, {\em inter alia}, this Bell quote from 1981~\cite{Bel81},   
``My own first paper on this subject starts with a summary of the EPR argument from locality to deterministic hidden variables.''  
While the concessions Norsen has made in Ref.~\cite{Nor15} thus eliminate some 
points of disagreement, there remain 
significant areas of dissension, related to Einstein's published views, 
to which we \blk turn in the following two sections. 
 
\section{``You keep using that word. I do not think it means what you think it means.''} 
\label{sec:locality}
Norsen, like most members of the realist camp, persistently uses the word `locality' as if everybody has 
always used it, and as if it is obvious that it means what  we   (following Bell's later usage) call ``local causality''. 
Neither of these conditions are true.  
To the best of our knowledge (which is admittedly limited) 
the word `locality' did not enter the physics lexicon until Bell's 1964 paper. EPR~\cite{EinPodRos35} certainly never used the word. 
Einstein, again to the best of our knowledge, did not use the word elsewhere. It had no generally accepted  
meaning until Bell gave it a meaning in 1964, and, as Norsen admits, that is not the meaning that he (Norsen) 
wants it to have.

Thus, when Norsen writes, in footnote 6 [our emphasis added], 
\begin{quote} \gold 
In the ``Autobiographical Notes'' Einstein gives a very clear argument, based on {\em locality}, \ldots . 
And of course the EPR paper provides an argument, again based on {\em locality} \ldots ,
\end{quote}
and similarly in other places in his paper, he \blk  
ignore essential differences between the localistic notions EPR used in 1935, those Einstein used in 1949, 
the notion of `locality' Bell introduced in 1964, and the notion of ``local causality'' which he introduced in 1976, 
and which Norsen wants to retroactively apply to all previous localistic notions of Einstein and Bell, 
albeit under the name of `locality'. 

In 1984 Jarrett~\cite{Jar84} adopted 
Bell's term `locality' with Bell's original meaning\footnote{It must be 
admitted, as Norsen does (4th-last paragraph of his Sec.~3) and as I do 
(in the final paragraph of Sec.~6.2, 
albeit contradicting a too-strong claim in the penultimate paragraph of Sec.~5), that Bell does not give a
mathematically precise formulation of `locality' in the way that Jarrett, for instance, does. 
 However, none of the plausible readings of Bell's words 
(neither mine, nor Norsen's) suffice to allow a derivation of predetermination from predictability. 
Thus, Norsen's raising of this issue does not help his case at all. The precise definition of `locality' given 
by Jarrett~\cite{Jar84} is the most general possible interpretation of Bell's words, which are set 
out for the reader's convenience later in this section. In particular, this notion  \blk is 
general enough to apply to probabilistic theories --- unlike Norsen's suggestion in his Eq. (3) --- which  is what Bell clearly 
intended since he appeals to his notion (albeit erroneously) \blk in his EPR paragraph prior to making the assumption of 
 determinism. Thus, it would seem churlish not to allow 
Bell the credit for this general concept.}, although immediately 
following this, Shimony proposed the new, rather awkward, term 
``parameter independence'' (PI) for the same notion~\cite{Shi84}, 
and this has since become well-recognized. (For the benefit of 
readers who are not familiar with these concepts, 
 we note here that it is because orthodox quantum mechanics 
respects Bell's 1964 notion of locality that operationalists 
readily accept Bell's 1964 theorem, as they see locality as the 
obvious assumption to keep and determinism 
as the obvious assumption to drop.)

Localistic assumptions are used in a number of places in the EPR argument, but never with any name, 
and sometimes without explicit acknowledgement. It is a subtle business to tease apart where they are making 
these assumptions in order for the formal argument they constructed in 1935 to work; see Ref.~\cite{Wis13} 
and Appendix~A of my paper. Despite the title of Bell's paper (an issue Norsen raises in the 2nd paragraph of his Sec.~3), 
there is no reason to think that Bell followed their formal argument; see Sec.~3.3 of 
my paper.  What Bell apparently had read was Einstein's 1949 notes, which are actually clearer on this score, 
introducing the localistic assumption that Bell quotes, that \grn``The real factual situation of system $S_2$ is independent of what is done with the system $S_1$, which is spatially separated from the former.'' \blk  Since Einstein does not name this notion, I have called it 
`no-telepathy' for reasons explained in my paper (Sec.~3.3; see also the next section).  

Einstein's `no-telepathy' assumption is not the same as the assumption of locality that Bell states 
and uses in 1964. Again, I have explained this in my paper; see also the following section. 
So that the reader may judge for him- or herself, here are all the relevant quotes from Bell's paper, in order, that 
either refer to Einstein's 1949 paper, or contain the word `locality', or repeat the notion that Bell defines 
(in the second item below) as `locality': 
\blu 
\begin{enumerate}
\item Einstein, Podolsky and Rosen \ldots\ \blk [gave] \blu an argument that quantum mechanics \ldots\ should be supplemented by additional variables \ldots\ to restore to the theory causality and locality [2].
\item It is the requirement of locality, or more precisely that the result of a measurement on one system be unaffected by operations on a distant system with which it has interacted in the past, that creates the essential difficulty.
\item Now we make the hypothesis [2], and it seems one at least worth considering, that if the two measurements are made at places remote from one another the orientation of one magnet does not influence the result obtained with the other.
\item The vital assumption [2] is that the result $B$ for particle 2 does not depend on the setting $a$, of the magnet
for particle 1, nor $A$ on $b$.
\item \blk [the violation of locality:] \blu the setting of one measuring device can influence the reading of another instrument, however remote.
\item Ref.~2: \grn ``But on one supposition we should, in my opinion, absolutely hold fast: the real factual situation of the system $S_2$ is independent of what is done with the \grn system $S_1$, which is spatially separated from the former.'' \blu A. ElNSTElN in Albert Einstein, Philosopher Scientist, (Edited by P. A. SCHILP) p.~85, Library of Living Philosophers, Evanston, Illinois (1949).
\end{enumerate}
\blk
The conclusion that Norsen draws from these quotes is that we should ignore Bell's own words, and 
understand \gold ``the passage from Einstein that Bell specifically chose to cite [three times] as, evidently, capturing his (Bell's) own understanding of this concept [of locality].'' \blk Norsen's 
conclusion seems unwarranted to us.  Bell never says, 
for example, ``Now we make the locality hypothesis of Einstein [2].'' 
The closest he comes is the sentence (iii) above. But there  
``[2]'' is just an interruption, in the same way that \blu ``and it seems one at least worth considering'' \blk is an interruption, of Bell's statement 
of his hypothesis. In this, and every other, instance, the reference ``[2]'' could be omitted from the sentence, and it would actually improve the grammatical and scientific clarity of the sentence. Thus, the conclusion 
we draw is that 
Bell gives his own definition of `locality', and uses the referencing of Einstein as an appeal to authority 
to justify the reasonableness of making an assumption like this.  Bell may well have thought that the notion 
he defined was the same as that which Einstein had enunciated, but, if so, he was wrong. 

Once again, Bell's 1971 paper~\cite{Bel71} supports our reading rather than Norsen's. 
If the quote from Einstein's 1949 paper~\cite{Ein49} really was the crucial phrase, 
which was what Bell meant by locality, and the starting point for his whole 
argument (from predictability to determinism to a contradiction with quantum mechanical predictions) 
then one would expect him to have reiterated this in his subsequent paper. But no. Not only does 
Bell not quote from Ref.~\cite{Ein49} in 1971, he does not even reference it. Nor does he adopt any 
terms, such as ``real factual situation'',  from that quote.

\section{What Einstein did}

The realist camp understandably prefers Einstein's 1949 argument with its explicit, if informal, localistic assumptions, 
over the formal and convoluted EPR argument with its oft-hidden localistic assumptions. However, in his recent paper,  
Norsen misrepresents both Einstein's 1949 argument, 
and my analysis of it (Sec.~3.3 of my paper). 

{\bf First,} regarding the passage where Einstein articulates what I called his 
`no-telepathy' assumption (see above), Norsen  asks (4th paragraph of his Sec.~4) 
\gold ``Does Wiseman \ldots\ mean to suggest that this passage also expresses Parameter Independence?'' \blk  
The answer is no! I explicitly say 
that Einstein's notion
 \red ``is different \blk [from Bell's notion of locality, \ie~PI] \red in that it requires not that Bob's {\em result} 
 $B$ be independent of Alice's setting $a$, but rather that the  
\grn``real factual situation'' \red of Bob's system be thus independent.'' \blk Even though Norsen gives 
(later in his Sec.~4) this quote from my paper, he remains, for reasons best known to himself, 
\gold ``suspicious'' \blk that I was 
\gold ``trying to suggest that Einstein, too, should be interpreted as having meant Parameter Independence.'' \blk To the contrary, the difference between Einstein's notion and PI is crucial for understanding why Einstein was right, at least under  
the realist interpretation of the wavefunction, in concluding that orthodox quantum mechanics 
involves `telepathic' influences (as explained in my paper and in the {\bf Fourth} point below). 

{\bf Second,} Norsen repeatedly implies that Einstein's notion here (which Norsen calls `locality' of course) 
will serve Bell's purpose in the EPR argument,  unlike Bell's notion (which Norsen also calls `locality'). 
This is incorrect and Einstein explicitly states that another assumption is necessary, which Bell does not 
mention, and which Norsen ignores in the main text of his paper~\footnote{That Einstein used two assumptions 
was stressed by Howard~\cite{How85}, who used the term `locality' for what I called Einstein's no-telepathy assumption. 
We think it better not to put that word (`locality') in Einstein's mouth, so to speak, because it has been so over-laden with meaning 
since Bell first introduced it (even by Bell himself, as we have seen).}.  
Norsen does this despite including (at the end of the final Einstein block-quotation in his Sec.~4) this telling Einstein quote: \grn
\begin{quote}
``One can escape from this conclusion \blk [that statistical quantum theory is incomplete] \grn only by either assuming that the measurement of $S_1$ (telepathically) changes the real situation of $S_2$ or by denying independent real situations as such to things which are spatially separated from each other. Both alternatives appear to me entirely\footnote{I must note an error in my reproduction of this quote in Sec.~3.3 of my paper, resulting from manual copying many years ago: I had `equally' in place of `entirely' here.} unacceptable.''\blk
\end{quote}\blk
As I said in my paper  
\red \begin{quote}``Einstein never states what qualifies something to be a  \grn ``real factual situation''. \red 
With some thought, one can propose a strong but reasonable formalisation for this concept 
such that the assumption (which, prior to the passage quoted \blk [by Bell]\red, Einstein makes explicitly) that systems have 
real factual situations, plus the no-telepathy supposition, has the same 
force as local causality \ldots. However, there is no suggestion of such 
a formalisation in Ref.~\blk\cite{Ein49}
\red.'' 
\end{quote}
\blk Norsen (following the final block-quotation of me in his Sec.~4) quotes only \blk a fragment of the above passage from my paper, thereby grossly misrepresenting what I say. 
Having given the Einstein quote above, he immediately examines the last sentence, but somehow misses the phrase \grn ``Both alternatives'' \blk in it: \gold
\begin{quote}
``What Einstein describes as \grn `unacceptable' \gold is unacceptable
precisely in the sense of violating the notion of locality
that he has articulated previously (in the sentence
partially quoted by Bell).''
\end{quote}
\blk Even when Norsen does, in his footnote~6,  mention the first of Einstein's alternatives, 
he claims that Einstein used the phrase \grn 
``entirely unacceptable'' \blk to describe the two assumptions \gold `jointly'\blk,  which is 
another clear misrepresentation. His footnote also suffers from the prevailing problem of using 
the term `locality' indiscriminately for  all localistic notions, as discussed in Sec.~\ref{sec:locality} above. 
\blk 

{\bf Third,} Norsen disputes my claim (which shortly follows the quote from my paper above) that 
Einstein, in 1949, did not use his two assumptions to \red ``make the argument that Bell wants to make, 
from predictability to determinism''. \blk Norsen says 
(in the paragraph containing the final block-quotation of me in his Sec.~4)  \gold ``[A]s is plain from passages from Einstein's essay that I \blk [Norsen] \gold 
have quoted above, this \blk [Wiseman's claim] \gold 
is at best misleading.'' \blk
Norsen is evidently using `plain' here in a sense we are unfamiliar with, 
since Einstein says nothing about the first party being able to predict the results of a measurement performed
by the second party. Indeed, Norsen includes a long footnote trying vainly to establish 
what is supposed to be `plain'. \blk 

Einstein's whole discussion is actually framed in terms of the conditional state of the second party. He says: 
\grn ``it appears to me that one may speak of the real factual situation of the partial system $S_2$.'' \blk 
(Note the tentative start to the sentence, consistent with this being exactly the assumption that he says later 
could be denied.) He then assumes (reasonably) that if quantum mechanics is complete, this real factual situation must
correspond to the conditional wavefunction of $S_2$. Assuming also `no-telepathy', Einstein 
obtains his contradiction. There is no mention of the second party making measurements at all. 
Certainly one could undertake to formalise Einstein's assumptions so as to make the 
argument from predictability to determinism. But equally certainly Einstein did not do this.

{\bf Fourth,} and finally for the purposes of the present paper, Norsen has this to say 
(in the paragraph preceding the final block-quotation of me in his Sec.~4): \gold 
\begin{quote}
Wiseman's suggestion that orthodox quantum mechanics
is some kind of \red ``counter-example'' \gold to Einstein's
argument (which Bell means to be summarizing) also underscores
the implausibly creative nature of Wiseman's
interpretation. Einstein's entire several-page-long discussion
(quoted above) is fundamentally about orthodox
quantum mechanics \ldots. 
The idea that Einstein somehow made an argument
for locally pre-determined values, but without bothering
to consider the concrete example of orthodox quantum
mechanics, is simply preposterous. His whole argument
is embedded in a discussion of orthodox quantum mechanics
from the very beginning.
\end{quote} \blk
This is probably the most important issue to address in terms of  
physical understanding (rather than who said what, why they said it, 
and whether they were right, in publications spanning the last 80 years.) I addressed 
it already in Sec.~3.3 of my recent paper, but Norsen ignores that. 

To begin, it is important to recognise that the phrase ``orthodox quantum mechanics'' is ambiguous. 
It could mean quantum mechanics which treats the wavefunction as something real, 
which undergoes collapse upon observation. This was the view promoted by von Neumann, 
and still widely held. 
On the other hand there is  the view of Bohr and (quoted here) Heisenberg:  
``In the Copenhagen interpretation of quantum mechanics, the objective reality has evaporated, 
and quantum mechanics does not represent particles, but rather, our knowledge \ldots\  of particles''~\cite{Hei58}.  
That is, the strictly operational view that there is no quantum reality to be represented by the wavefunction or anything else.

Einstein's argument certainly establishes that the realist interpretation 
of the wavefunction relies upon `telepathic' change in the real situation of $S_2$. Unfortunately for Bell,  
his 1964  `locality' notion, which is weaker than Einstein's 1949 `no-telepathy' notion,  
is not violated even by remote wavefunction collapse. The reason is that Bell (in 1964) 
only ever mentions \blk the effect of the measurement choice by the first party (Alice) on the {\em outcome} obtained 
by the second party (Bob). In orthodox quantum mechanics, where there are 
no hidden variables $\lambda$ (beyond perhaps the wavefunction $\psi$ if 
the preparation is mixed) Alice's measurement choice has 
no statistical relation with Bob's outcome. This is presumably not what 
Bell thought or intended, but we cannot rewrite history to have him 
define `locality' differently from how he did define it. 

Einstein's argument does not apply, however, 
 to the Copenhagen interpretation\footnote{At least the Copenhagen interpretation in the post-EPR era; see Ref.~\cite{Beller99}}, precisely because that 
interpretation embraces \grn ``denying independent real situations as such to things which are spatially separated from each other.'' \blk 
Of course the Copenhagen interpretation does not deny \grn ``real situations as such'' \blk only 
in the context of spatially separated things that are entangled. It denies them to all quantum systems. 
As already discussed above, and as the above quote will remind the reader, Einstein assuredly did not 
present his argument \gold ``without bothering
to consider the concrete example of orthodox quantum
mechanics,'' \blk even if we understand that to be the theory as promoted by Bohr and Heisenberg. 
Rather he explicitly acknowledged the possibilities favoured by his colleagues, even when he found 
them \grn`unacceptable'\blk. 
Norsen could learn something from this. 

\section{The bigger picture}

Having addressed \blk Norsen's diverse criticisms, and clarified how Bell's 
and Einstein's notions apply to ``orthodox quantum theory'' in its different 
interpretations, it is useful to step back and ask the question: why does this 
issue raise such passions? 

Norsen concludes his paper by saying that by \gold 
``telling the `operationalists' that they were right all along, in how they understood Bell's 1964 paper,'' \blk  
I was \gold ``doing a great disservice to Bell on this 50th anniversary of his great achievement.'' \blk 
It certainly gave me no pleasure, on the anniversary of Bell's undeniably great achievement, to be 
uncharitably probing the meaning of individual words and phrases in a ground-breaking paper 
which was written quite quickly, taking only a few months from conception to completion. 
But it is a job that desperately needed doing, in my opinion, to show the realist camp 
why the operationalist reading of Bell's 1964 paper is ``the one almost universally reported''~\cite{Bel81}. 
This last phrase reminds us, however, that Bell in later life preferred to present his 1964 
paper in a different way, as merely a less developed version of his 1976 theorem, which 
is favoured by realists. 

This, we \blk think, is what Norsen really objects to: that I argued for the existence of 
a Bell's theorem (the 1964 one) which is acceptable to operationalists because it allows 
them to keep a localistic notion (`locality') while accepting the violation of Bell's second assumption, 
determinism (or, as Bell actually calls it, predetermination). This is supposedly a disservice to Bell  
because in later life he rejected (implicitly) the notion of locality he had articulated in 
1964, and promoted the notion of local causality instead. The 1976 theorem fits with Bell's (and Norsen's) 
realist sympathies because it requires no second assumption along the lines of determinism; 
it requires even operationalists to accept the violation of local causality.  
But what realists often fail to realise is that an operationalist may accept the violation of local causality
in this manner: ``Ok, it is violated, but that doesn't interest me much \blk because local causality is not how 
I understand Einstein's principle of relativity. `Locality' as Bell 
defined it in 1964 seems to me a much better expression of it. \blk So I am much more interested in Bell's 1964 theorem.'' 

How would matters change if we were to admit Norsen's central point? If we were to say, 
yes, Bell's 1964 theorem is the same as his 1976 theorem, because Bell always did mean 
the same thing by `locality' and ``local causality''. Bell's theorem (no need to distinguish 
the years anymore) would then be comparatively \blk 
uninteresting to operationalists. However, there would 
still be a valid theorem 
starting not from the assumption of (let us call it for clarity) local causality, but from the 
joint assumptions of locality (as we \blk will continue to call it) and 
predetermination. That theorem, which would not be called Bell's theorem any more, 
would still appeal to operationalists. And so operationalists would celebrate this 
important theorem of some other quantum physicist(s), rather than Bell --- perhaps 
CHSH~\cite{CHSH69} since they do not present the EPR argument\footnote{Indeed, they appear skeptical of it, 
saying only 
that EPR's ``paradox \ldots led them to infer that quantum mechanics is not a complete theory'', 
explaining what that means, and then summarizing 
Bohr's argument (as they understand it)  against the validity of EPR's conclusions.} 
prior to setting out their assumptions of deterministic outcomes and locality (as they do indeed 
call it, following Bell). Would that be doing a service to Bell?

Perhaps Norsen thinks it would be a service to Bell, \blk because, as far as he is concerned, operationalists 
inhabit an intellectual wilderness, isolated from the mainstream of physics which has 
always sought to understand the reality behind appearances, and best kept isolated. 
Norsen also might argue that, logically, operationalists should not be interested in the 
``locality plus predetermination'' theorem because locality is not a strictly operational 
concept. The strictly operational concept is signal locality, and actually there is a 
``signal locality plus predictability'' version of Bell's \blk theorem of some importance in quantum information~\cite{Mas06}
as well as quantum foundations~\cite{CavWis12}. This is a fair point, 
but most operationalists are not so strict. For example, they are willing to entertain 
``hidden variables'' in the form of an ensemble of pure quantum states $\psi$ that 
may underlie a mixed preparation $\rho$. Also, they often talk of 
$\psi$ as a real thing undergoing measurement-induced collapse. \blk In that context, non-strict operationalists 
would still be glad to know that these versions of orthodox quantum theory respect \blk 
not just signal locality but also locality, as discussed above. 

So, while we can see that staunch realists could see my analysis of Bell's 1964 paper 
as doing him a disservice, for everybody else  
my mild criticisms of Bell would surely seem less of a disservice to him 
than what Norsen would have us do: (i) deny him credit for the 1964 
theorem (as we see it) favoured by operationalists; and (ii) turn his 1964 theorem into a non-theorem 
by claiming that it rests on a single assumption (locality), which as defined in Bell's paper 
is manifestly inadequate for deriving Bell's result.  
\blk

\newpage 

\section*{Acknowledgements} 
Thanks to the anonymous referee, Matt Pusey, Bryan O'Gorman, 
and Travis Norsen for helpful comments. 
This research was supported by the ARC Discovery Project DP140100648. 

\section*{References}
\bibliography{../../QMCrefsPLUS}

\end{document}